%% file: paper_jlt_PDFincludes.tex
\documentclass[journal]{IEEEtran}

\usepackage{paper_fehenberger}
\input{packages_paper}
\newboolean{makePlotsBlackWhite} 
\setboolean{makePlotsBlackWhite}{false} 

\input{myAcronyms}

\begin{document}
%
\title{Mutual Information as a Figure of Merit\\for Optical Fiber Systems}

\author{Tobias~Fehenberger,~\IEEEmembership{Student Member,~IEEE}
        and~Norbert~Hanik,~\IEEEmembership{Senior Member,~IEEE}
\thanks{T. Fehenberger and N. Hanik are with the Institute for Communications Engineering, Technische Universität München, 80333 Munich, Germany (\mbox{E-mail:}~tobias.fehenberger@tum.de, norbert.hanik@tum.de).}
}

%


\maketitle


\begin{abstract}
Advanced channel decoders rely on soft-decision decoder inputs for which
\MI is the natural figure of merit. In this paper, we analyze an optical fiber system by evaluating \MI as the maximum achievable rate of transmission of such a system.
%
%
\MI is estimated by means of histograms for which the correct bin number is determined in a blind way.
The \MI estimate obtained this way shows excellent accuracy in comparison with the true \MI of $16$-state \QAM over an additive white Gaussian noise channel with additional phase noise, which is a simplified model of a nonlinear optical fiber channel. We thereby justify to use the \MI estimation method to accurately estimate the \MI of an optical fiber system. In the second part of this work, a transoceanic fiber system with $6000$~km of standard single-mode fiber is simulated and its \MI determined. Among rectangular QAMs, $16$-\QAM is found to be the optimal modulation scheme for this link as to performance {in terms of }\MI and {requirements on components and digital signal processing}. For the reported \MI of $3.1$~bits/symbol, a minimum coding overhead of $29$\% is required when the channel memory is not taken into account. By employing ideal single-channel digital back-propagation, an increase in \MI by $0.25$~bits/symbol and $0.28$~bits/symbol is reported for $16$-\QAM and $64$-QAM, respectively, lowering the required overhead to $19$\% and $16$\%.
When the channel spacing is decreased to be close to the Nyquist rate, the dual-polarization spectral efficiency is $5.7$~bits/s/Hz, an increase of more than $2$~bits/symbol compared to a $50$~GHz spacing. 
\end{abstract}


\begin{IEEEkeywords}
Mutual information, achievable rate, channel coding, long-haul fiber communication, digital back-propagation
\end{IEEEkeywords}
%
\IEEEpeerreviewmaketitle

\acresetall
\section{Introduction}
\IEEEPARstart{T}{he} demand for increased data rates in optical long-haul communications has been growing for several years. A capacity crunch \cite{Chraplyvy2013} seems inevitable and spatial multiplexing is considered by many as the only solution to this problem because it offers lower costs due to integration of components \cite{Winzer2013}. The investment costs to deploy new fibers, however, are high and spatially multiplexed systems have not reached a commercial level for long-haul fiber communication. Hence, optical networks operators rely for the time being on currently deployed single-mode fibers instead of switching to new fiber technologies.
For those single-mode fibers, an obvious improvement on the de-facto standard long-haul $100$~Gbit/s/channel systems operating with \QPSK at $28$ billion symbols per second per polarization can be made by increasing the order of the modulation scheme. The accompanying increased susceptibility to noise is in conflict to the requirement that the \BER after decoding (denoted \BERout) must be kept at a constant level of around $10\textsuperscript{-15}$ and that the transmission distance must remain the same. The required protection of bits can be achieved on the information-theoretic layer by applying more advanced coding schemes. Most of these schemes have a decoder that operates with soft inputs, offering an improved decoding threshold because the reliability of a bit decision is taken into account. As we will outline in detail in Section~\ref{sec:inftheory}, using \MI as a figure of merit is more natural to this family of soft-decision decoders. More importantly for this work, it also lets us determine the maximum achievable rate of an optical fiber channel.

\MI has been a key quantity in the field of information theory since Shannon's seminal paper \cite{Shannon1948}. {In optical communications, it has been used to state lower-bound estimates on channel capacity} \cite{Mitra2001}, \cite{Essiambre2010}. Recently, \MI has drawn attention as an \FEC decoding threshold. Optical transmission experiments \cite{Leven2011} suggest that \MI is more reliable than the \Q factor in order to estimate the post-decoding \BER of a particular \SD decoder. \MI predicting the \BER after decoding has also been applied in simulations to long-haul links \cite{Fehenberger2013}. It is also used as decoding threshold in record experiments \cite{Salsi2013}. When \MI represents a decoding threshold, {it is calculated from bit-wise log-likelihood ratios (LLRs) representing the soft a-priori information coming from the channel to the decoder. To calculate LLRs, a distribution of the received symbols must be assumed. Typically, the noise is taken to be Gaussian, although this assumption does not always hold in optical communications} \cite{Cho2012}.

In this work we do not make any assumptions on the noise characteristics of an optical channel. Instead, \MI is estimated by means of histograms for which the correct number of bins is determined with a blind and accurate method. 
When applied to an optical fiber link, \MI is used to find the maximum potential data throughput over a fiber. Further, \MI serves as a relative figure of merit to assess the impact of changing a design parameter or adding a routine to the receiver with respect to achievable rate. This analysis is performed exemplarily by employing \DBP and for varying spectral channel spacings.

%
%
%
This paper is organized as follows. Section~\ref{sec:inftheory} explains information-theoretic fundamentals of channel coding and introduces a reference channel model. In Section~\ref{sec:estimatingMI}, a method for \MI estimation is presented and its accuracy verified. Simulating an optical transmission system and applying \MI to it is outlined in Section~\ref{sec:simulations}. Section~\ref{sec:conclusion} concludes this work.
\newpage
%
\section{Information Theory}\label{sec:inftheory}
\subsection{Hard and Soft Decisions}\label{subsec:hd_sd}%
Every code consists of at least two entities, an encoder that maps bits to codewords and a corresponding decoder that tries to undo this mapping when only a distorted copy of the sent codeword is available at the decoder. The input into such a decoder can be \HD or \SD, depending on the design of the symbol demapper and the code. For simplicity, we will refer to codes whose decoders use soft (hard) input as \SD (\HD) codes. Note that there are code families that work with both hard and soft decoding. See for example \cite{Koetter2003} on how \SD decoding is applied to \RS codes.

For hard decisions, the binary sequence after the demapper is quantized such that all information available at the decoder input is binary. In contrast, \SD decoding operates with more than two quantization levels. The reliability of a received symbol with respect to the decoder operation is increased by taking into account the distance of a symbol to its decision boundaries. The closer a symbol is to the decision region of a different symbol, the less reliable the decision on this symbol is going to be. With an infinite number of quantization levels, it can be shown \cite[Sec.~6.8]{Proakis} that \SD decoding outperforms \HD codes by up to $2$~dB of energy per bit for the code rate $R\rightarrow 0$. For a practical rate up to about $0.9$, the potential gain is still larger than $1$~dB. Hence, even the most powerful \HD codes are outperformed by \SD codes.

For many years, \HD codes such as the ubiquitous \RS code with $K=8\cdot239$ information bits per block of length $N=8\cdot255$ bits \cite{ITUG975} have been the most prominent coding schemes in optical long-haul systems. {Improved} \FEC for optical communications \cite{ITUG9751} is mostly based on hard decisions as well. The focus on \HD codes can be mainly attributed to two of their characteristics. Firstly the computational complexity is small because computations are made with integers rather than floating points. Additionally, it is possible to exactly determine the decoding performance by solely considering the \BER at the decoder input (denoted \BERin) when perfect interleaving is assumed. For the \RS($255, 239$) code of rate $R=K/N=239/255$ (which corresponds to a coding overhead of $(1/R)-1 \simeq 6.7\%$), the \BERin must be around $8.3\cdot10\textsuperscript{-5}$ in order to reach a \BERout of $10\textsuperscript{-15}$ { }\cite{ITUG975}.
{Further improvements on} \HD code design allow the \BERin to go as high as $4.64\cdot10\textsuperscript{-3}$ \cite{Smith2012} for a coding overhead of $6.7\%$ and a \BERout of less than $10\textsuperscript{-15}$.
This so-called \ac{FEC} decoding threshold is usually stated in terms of the \Q~factor, which is defined as
\begin{equation}\label{eq:Q2Factor}
 Q^2 = 20\log_{10}(\sqrt{2} \cdot \text{erfc}^{-1}(2 \cdot \BERin)) \quad \text{in dB},
\end{equation}
where $\text{erfc}^{-1}$ denotes the inverse of the complementary error function.
\Q is thus a perfectly suitable figure of merit for an \HD code if the \BERout is to be examined.
Unfortunately, this one-to-one correspondence between \BERin (or equivalently \Q ) and \BERout is not given for most \SD codes. It is in general impossible to analytically determine the decoding performance of an \SD code with finite block-length by considering an input at the decoder. 
For most \SD coding schemes, the decoding operation starts with soft a-priori information from the channel, usually in the form of LLRs. \BERin does not play a role at the decoder input and it consequently makes no sense to cling to \Q as a figure of merit for these codes. Instead of using \Q as characteristic prediction for \BERout, \MI is the more natural and also more robust figure of merit \cite{Leven2011}.

In addition, \MI enables us to evaluate the influence of certain design parameters in terms of net achievable rate. We will illustrate this with a short example. Consider a nonlinear, \WDM optical communication link with either an \HD or an \SD coding scheme in place. For this system the influence of increasing the \WDM channel spacing is to be examined. We expect the signal quality in general to improve for an increased spacing because the impact of nonlinear signal-signal interactions is weaker. For the \HD code, the improvement from increasing the spectral spacing translates to an increased \Q and consequently a smaller \BERout that can be calculated analytically. If the required \BERout is already reached, the improvement due to larger \WDM spacing essentially corresponds to an assured gain in link budget. In contrast, a gain in \Q does not translate to any characteristic of an \SD decoder. Most importantly, not even a lower \BERout can be guaranteed.
%
%
%
Let us now consider the \SD code and \MI as figure of merit at its decoder input. Similarly to the previous \HD case, \MI grows with increasing channel spacing.
Let us also assume a capacity-achieving code with perfectly flexible coding rate. These assumptions are fulfilled by spatially-coupled \LDPC codes \cite{Kudekar2013} and polar codes \cite{Arikan2009}. Both coding schemes provably achieve the capacity of an \AWGN channel while maintaining low decoding complexity. Polar codes further offer a perfectly flexible rate by using the $K$ best polarized sub-channels. Exploiting these properties, we can use the gain in \MI to lower the rate of the used \SD code such that we operate right at the achievable rate. A change in \MI hence directly translates to change in net available data rate. In the following, we elaborate on this concept. We also show simulation results for the presented example in Section~\ref{subsec:simresults}.
%
%
%
\subsection{Channel Capacity, Mutual Information, Achievable Rate}\label{subsec:mi}
Consider two random variables $X$ and $Y$ representing the channel input and output, respectively. Shannon \cite[Sec. 12]{Shannon1948} tells us that the capacity of a channel is the maximum rate of transmission, or equivalently, the \MI $I(X;Y)$ maximized over all possible input distributions $p_X$,
\begin{equation}\label{eq:capacity}
 C = \max_{p_X} I(X;Y).
\end{equation}
The \MI $I(X;Y)$ represents the amount of information about $X$ that is contained in $Y$ when $X$ is transmitted. It is also the achievable rate, i.e. the rate for which the \BER tends to zero as the block length of the code goes to infinity. Rates that are smaller than the channel capacity yield an arbitrarily small \BER when the block length is sufficiently large \cite[Sec. 7.6]{Cover2006}.\\
For practical applications, restrictions are made to the channel input and output. The channel input $X$ is fixed to a certain modulation format, which is required to map information onto the symbols at the transmitter. Hence, $X$ is a fixed discrete random variable with alphabet $\mathcal{X}$ whose cardinality $|\mathcal{X}|$ is the modulation order $M$. In this work, we further assume the input distribution ${p_X}$ to be uniformly distributed. The channel output $Y$ is also discrete, which is motivated by using an \ADC at the receiver. Under these assumptions, the \MI is (see for example \cite[Sec.~23.2]{Hanzo2004})
\begin{equation}\label{eq:MI}
I(X;Y)=\sum \limits_{m=1}^{M} \sum \limits_{\mathcal{Y}} p_X(x_m) p_{Y|X}(y|x_m) \log_2 \left[\frac{p_{Y|X}(y|x_m)}{p_Y(y)}\right]\!\!,
\end{equation}
where $p_{Y|X}(y|x_m)$ is a conditional \pmf of an output realization $y \in \mathcal{Y}$ given the $m$\textsuperscript{th} input realization $x_m \in \mathcal{X}$ and $p_Y(y)$ is the \pmf of the output. The second sum represents summing over all values of $\mathcal{Y}$ in the complex domain. For a uniformly distributed input, $p_X(x)=\frac{1}{M}$.

An important aspect of the optical channel is its inherent memory. In this work, we neglect any memory and do not make use of the statistical relations between neighboring symbols. Thereby, a memoryless auxiliary channel is introduced. The \MI of this auxiliary channel is proven to be a lower bound of the \MI of the true channel with memory \cite[Sec. VI]{Arnold2006}. This means that the \MI of the auxiliary channel is still an achievable rate. This relation between \MI and achievable rate serves as basis for the analysis carried out in Section~\ref{sec:simulations}.\\
\begin{figure}[t]%
	\includegraphics[width=\figurewidth,height=\figureheight]{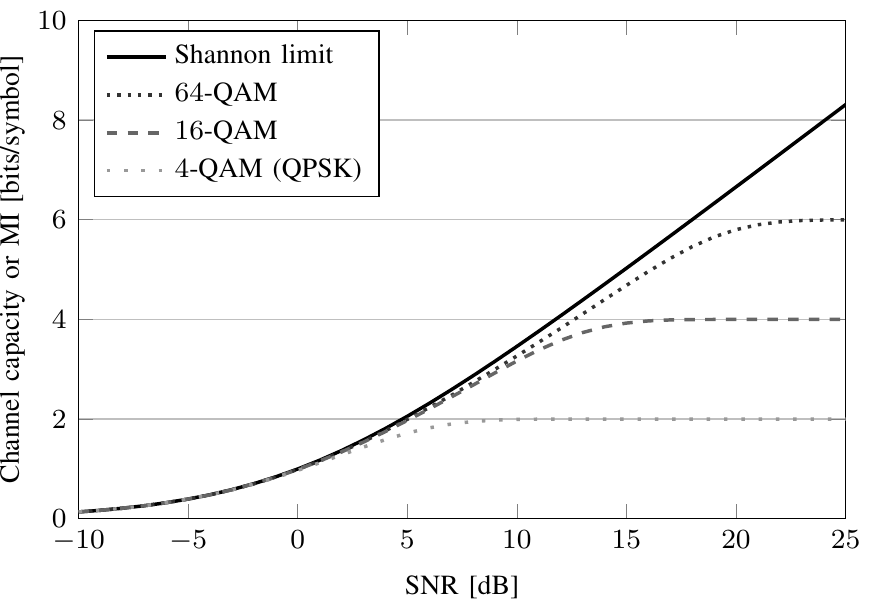}%
	\caption{{Circularly symmetric AWGN channel capacity (Shannon limit) and MI of QAM of order $4$, $16$, $64$ over this channel.}}%
	\label{fig:Capacities_AWGN}%
\end{figure}%
For a complex cs AWGN channel, {whose noise distributions along the two complex dimensions are independent}, the capacity-achieving distribution under an average-power constraint is continuous complex Gaussian. Its channel capacity is given by the well-known formula $C=\log_2(1+\textnormal{SNR})$. This quantity as well as the \MI of \QAM of respective order $M=\{4, 16, 64\}$ transmitted over this channel are shown in Fig.~\ref{fig:Capacities_AWGN} as a function of \SNR in dB.
Among the many important features to be seen in Fig.~\ref{fig:Capacities_AWGN}, we want to point out that the \MI of each discrete modulation format saturates at a maximum of $\log_2 M$~bits/symbol for high \ac{SNR}s. Also, the \MI for all depicted modulation schemes is approximately the same in the low-to-medium \SNR regime.%
\subsection{Reference Channel Model}\label{subsec:channelmodel}
Estimating the optical fiber capacity, or at least lower bounds thereof, has been a vivid research topic for several years \cite{Mitra2001} \cite{Essiambre2010}. So far, no explicit capacity expression is known as there is no analytical channel model for a dispersive nonlinear fiber channel with inline amplification. However, several approximate channel models exist that simplify the noise distributions of the optical channel by making certain assumptions. Among these models, the Gaussian noise model \cite{Poggiolini2012}, the \PCAWGN model \cite{Goebel2011}, \cite{Leoni2013} and an input-dependent complex AWGN model with non-diagonal covariance matrix \cite{Cho2012} have received notable attention. The latter model will not be considered further as it models the orientation of the received ellipse-shaped constellation symbols in dependence of the channel input, which is not suitable for our analysis. In this work, we use the \PCAWGN channel as reference. The time-discrete form of the \PCAWGN channel is \cite{Goebel2011}
\begin{equation} \label{eq:PCAWGN}
Y = X \cdot e^{j\Phi} + N,
\end{equation}
where the additive noise $N$ is a complex Gaussian-distributed random variable with zero mean and noise power spectral density $\frac{N_0}{2}$ per complex dimension, $N\sim\mathcal{N}(0,N_0)$. $\Phi$ models the phase noise as real-valued wrapped \AWGN with variance $\sigma^2_\Phi$. The well-known \AWGN channel is thus a special case of \eqref{eq:PCAWGN} with $\sigma^2_\Phi=0$. The \PCAWGN {channel is essentially a }\cs \AWGN {channel with additional phase noise such that in the absence of} \AWGN, {excessive phase noise leaves only the amplitude to transmit information. This makes the channel only \textit{partially} usable to convey information, hence its name} \cite{Goebel2011}. {As a simplified approximation to a nonlinear channel, the additional phase noise models nonlinearities due to the Kerr effect as well as residual linear impairments that are not fully compensated at the receiver. A higher optical transmit power resulting in more nonlinearities, an or insufficient linear compensation at the receiver lead to an increase in phase noise.} The \PCAWGN channel exhibits the typical concave function of channel capacity estimate over optical launch power, known from e.g. \cite{Mitra2001}, if for increasing signal power the phase noise is modeled such that it grows faster than the signal power. {The nonlinear regime in optical transmission is approximately represented by} $\sigma^2_\Phi$ less than $10^{-1}$ \cite{Leoni2013}.

The transition \pdf of the \PCAWGN channel \eqref{eq:PCAWGN} is obtained by combining the \pdf of a real-valued wrapped Gaussian random variable and the \pdf of complex \AWGN \cite{Leoni2013},
\begin{align} \label{eq:condPDFpcAWGN}
p_{Y|X}(y|x) = \int_{-\pi}^{+\pi} &\frac{1}{\pi N_0} e^{\frac{-|y-xe^{j\varphi}|^2}{N_0}}  \cdot \nonumber \\
& \cdot \frac{1}{\sqrt{2\pi\sigma_{\Phi}^2}} \sum_{k=-\infty}^{+\infty} e^{-\frac{(\varphi-2 \pi k)^2}{2\sigma_{\Phi}^2}} \textrm{d}\varphi.
\end{align}
This \pdf fully describes the relation between channel input $X$ and output $Y$. When the input $X$ is known at the receiver, the \MI $I(X;Y)$ of \eqref{eq:MI} can be calculated. This is done in Section~\ref{subsec:NumericalVerification} to obtain a \MI reference for the \MI estimate that is obtained with the method presented in the following.

\section{Estimating Mutual Information}\label{sec:estimatingMI}
If the \ac{pdf}s $p_{Y|X}(y|x_m)$ and $p_Y(y)$ of \eqref{eq:MI} are not known or cannot be stated explicitly for the channel of interest, \MI cannot be determined numerically but must be estimated. As \MI between random variables is such a fundamental quantity, its estimation is an important research topic not only in information theory but also in, e.g., genetics \cite{Daub2004} and neural computation \cite{Paninski2003}. Accordingly, a lot of known techniques to estimate \MI are known, such as B-spline functions \cite{Daub2004}, kernel density evolution \cite{Moon1995}, Gauss–Hermite quadratures \cite[Sec. III]{Alvarado2011}, second-order Taylor expansion \cite{Goebel2005}, nearest neighbor statistics \cite{Kraskov2004}, and histograms \cite{Darbellay1999}.
Among them, \MI estimation by means of histograms is considered in detail in this work as histograms are the conceptually most intuitive and also most straightforward technique to implement. A comparison of different techniques is beyond the scope of this work as the main focus is to find an accurate estimation technique and then apply this method to an optical fiber system.
\subsection{Histograms}\label{subsec:Histograms}
The principle of histograms is to divide the value range of the symbols $\mathbf{S}$ into a number of intervals called bins. The number of symbols falling into each bin represents the value of the discrete distribution when apt normalization is performed.\\
Correctly choosing the number of bins is crucial to obtain an estimate of the \ac{pdf}s and thus of \MI that is as exact as possible. For too few bins, the estimate becomes too coarse and does not accurately resemble the distribution. If the number of bins is chosen too large, \MI tends to be overestimated. This is because \MI estimation is based on a sequence of finite length. Using a large number of bins, there are bins into which no received symbols fall, especially at the low-probability tails of the distribution to be estimated. These tails are usually the portion of the continuous \pdf $p_{Y|X}(y|x_m)$ that overlaps into the decision region of a different sent symbol $x_{k \neq m}$. The corresponding summands of this overlap region can be negative for $\log_2 [\frac{p_{Y|X}(y|x_m)}{p_Y(y)}] < 0$ in \eqref{eq:MI}, and $I(X;Y)$ is decreased by them. If, however, no symbol falls into these bins, they must be excluded from the calculation as the $\log_2$ in \eqref{eq:MI} is not defined when either the {numerator} or the denominator equals zero. This means that terms that reduce the \MI are wrongfully left out if we do not take care to produce as few empty bins as possible while recreating the shape of the distribution as good as possible.

\begin{figure}[t]%
	\includegraphics[width=\figurewidth,height=\figureheight]{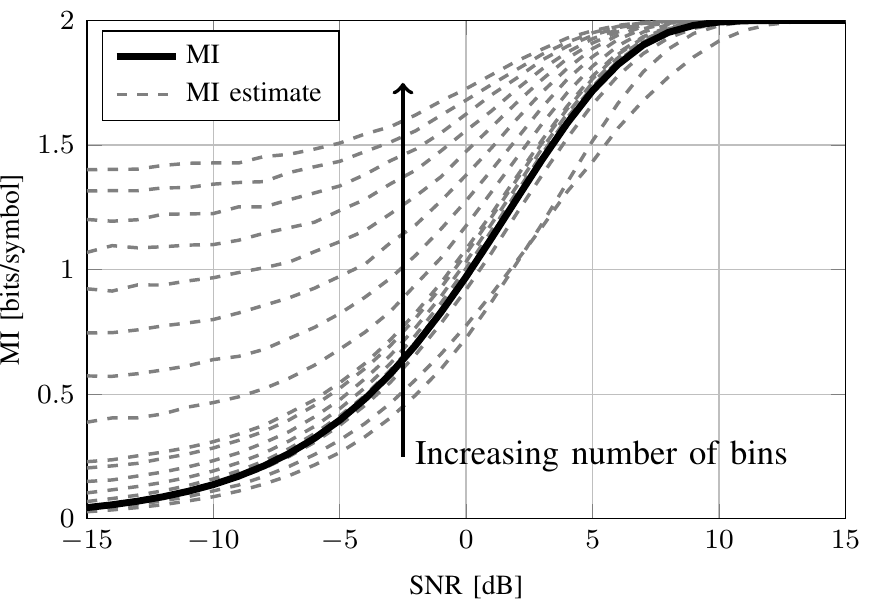}%
	\caption{{Numerically calculated MI} and \MI estimates obtained with histograms of different bin sizes for \QPSK over a \cs \AWGN channel. The direction of the arrow indicates increasing number of bins.}%
	\label{fig:MI_Knuth_QPSK_AWGN_IncreasingBinSize}%
\end{figure}%
To illustrate the effect of using an overly coarse or fine histogram, Fig.~\ref{fig:MI_Knuth_QPSK_AWGN_IncreasingBinSize} depicts the \MI over \SNR for $2\textsuperscript{14}$ \QPSK symbols transmitted over a \cs \AWGN channel. The \MI estimate is obtained with a joint histogram of in-phase and quadrature. The vertical arrow indicates increasing bin number, from $2$ bins up to $500$ bins per dimension. A too small number of bins leads to an incorrectly small \MI estimate. Contrarily, too many bins result in overestimating the actual \MI by far. Note that it seems as if there were a certain constant bin number that provided sufficient estimation accuracy. Correctly applying a histogram, however, depends not only on the \pdf shape but also on the number of symbols that are evaluated jointly. The more samples are considered for one particular instance of a histogram, the more likely it is that symbols fall into the low-probability tails of the \pmf and the histogram must be chosen finer. In the following, a method is presented that allows us to calculate the bin number for which we obtain an accurate MI estimation while taking into account the dependence on both the \pdf shape and the sequence length.
%
\subsection{Find the Correct Number of Bins}\label{subsec:MIestimate}
Finding the right number of bins for a histogram is a well-known task, see for example \cite{Darbellay1999}, \cite{Scott1979}, \cite{Shimazaki2007}, \cite{Knuth2013}. In this work, the input into the \MI estimation algorithm is two-dimensional. Also, we require the \pdf estimation to be blind, i.e. only the received symbols may have an impact of the \MI estimate. Any a-priori knowledge of the channel characteristics or choosing suitable algorithm parameters should not be required in order not to influence the results by making assumptions.

\cite{Scott1979} is proven to be optimal for one-dimensional normally distributed data, which does not match the properties of the communication system at hand. The method presented in \cite{Darbellay1999} is suitable for multi-dimensional data. However, certain parameters for the estimator function must be chosen in order for the algorithm to work, which makes the method not blind. For two-dimensional data, \cite{Shimazaki2007} optimizes over both dimensions but gives the same bin number for in-phase and quadrature. The resulting square bins are not optimal for noise that is not symmetrical as it is the case in highly nonlinear fiber optics. Although there might be many more promising techniques, the method presented in \cite{Knuth2013} is chosen because it is blind and also expandable to two dimensions in an easy manner. We will see in Section~\ref{subsec:NumericalVerification} that \MI estimation using this method is accurate for the \PCAWGN channel (our reference channel). Finding a potentially superior method is left for future work.

The method described in detail in \cite{Knuth2013} is based on the idea that a histogram is a piecewise-constant density model of the \pdf to be estimated. This allows to state the posterior probability of the bin numbers given a sequence of symbols $\mathbf{S}$ of length $N_\textnormal{s}$. The pair of bin numbers of the in-phase and quadrature component is found jointly over both dimensions and denoted as $(\hat{N}_{\textnormal{b},\mathcal{I}}, \hat{N}_{\textnormal{b},\mathcal{Q}})$. For the ease of notation, the product $N_{\textnormal{b},\mathcal{I}} \cdot N_{\textnormal{b},\mathcal{Q}}$ is written as $N^*_\textnormal{b}$. \cite{Knuth2013} states that
\begin{align}\label{eq:argmaxBins2D}
(\hat{N}_{\textnormal{b},\mathcal{I}}, \hat{N}_{\textnormal{b},\mathcal{Q}})\!=&\!\argmax_{N_{\textnormal{b},\mathcal{I}}, N_{\textnormal{b},\mathcal{Q}}} \bigg\{ N_\textnormal{s} \ln N^*_\textnormal{b}\!+\!\ln \Gamma(\frac{N^*_\textnormal{b}}{2})\!-\!N^*_\textnormal{b} \! \cdot \! K + \nonumber  \\ 
  & -\!\ln\!\Gamma (\!N_\textnormal{s}\!+\!\frac{N^*_\textnormal{b}}{2}\!)\!+\!\!\!\sum_{k=1}^{N_{\textnormal{b},\mathcal{I}}}\!\sum_{l=1}^{N_{\textnormal{b},\mathcal{Q}}}\!\ln\!\Gamma(\!n_{k,l}\!+\!\frac{1}{2}\!)\! \bigg\},
\end{align}
where $\Gamma(x)$ denotes the Gamma function defined as $\Gamma(x)=\int_0^\infty e^{-t} t^{x-1}\textnormal{d}t$ and $K=\ln \Gamma(\frac{1}{2})$ is a constant.
When a histogram operation with $(N_{\textnormal{b},\mathcal{I}}, N_{\textnormal{b},\mathcal{Q}})$ bins is applied to the symbols $\mathbf{S}$, 
the number of samples falling into the $k^\textnormal{th}$ and $l^\textnormal{th}$ bin along the real and imaginary dimension, respectively, is denoted $n_{k,l}$.

The maximization in \eqref{eq:argmaxBins2D} is solved numerically with a brute-force approach for $1$ to $N_\textnormal{b,max}$ bins per dimension. For a reasonably large $N_\textnormal{b,max}$, the results of Fig. \ref{fig:MI_Knuth_QPSK_AWGN_IncreasingBinSize}, the following simulation results and \cite{Knuth2013} suggest that an argument that maximizes \eqref{eq:argmaxBins2D} will not be found by probing more than $N_\textnormal{b,max}$ bins. This limits the computational time of finding the correct bin numbers.\\
For a block of symbols $\mathbf{S}$, the corresponding \MI is calculated at the receiver as follows.
\begin{enumerate}
 \item Determine $(\hat{N}_{\textnormal{b},\mathcal{I}}, \hat{N}_{\textnormal{b},\mathcal{Q}})$ according to \eqref{eq:argmaxBins2D}.
 \item Perform a two-dimensional histogram operation with $(\hat{N}_{\textnormal{b},\mathcal{I}}, \hat{N}_{\textnormal{b},\mathcal{Q}})$ bins on $\mathbf{S}$. Obtain $p_Y(y)$ by dividing the number of symbols falling into every bin by $N_\textnormal{s}$.
 \item Store the bin positions of step 2.
 \item Determine $p_{Y|X}(y|x_m)$ by performing another histogram with the bin positions stored in step 3 on those of the symbols $\mathbf{S}$ that belong to the $m$\textsuperscript{th} symbol. After normalization, $p_{Y|X}(y|x_m)$ is obtained.
 \item Repeat step 4 for every $m$ from $1$ to $M$.
 \item Calculate the \MI according to \eqref{eq:MI}. 
\end{enumerate}
\subsection{Numerical Verification}\label{subsec:NumericalVerification}
\begin{figure}[t]%
\includegraphics[width=\figurewidth,height=\figureheight]{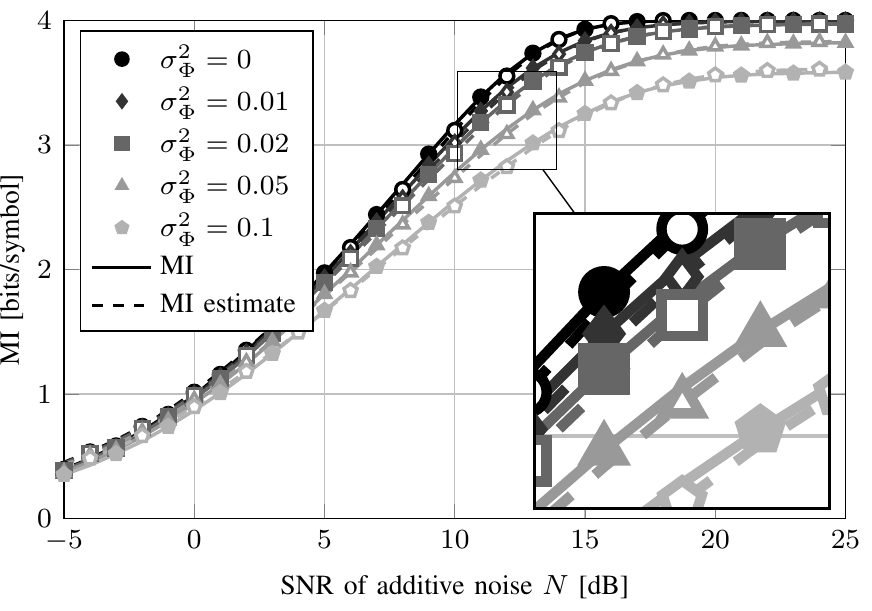}%
	\caption{{Numerically calculated MI} and \MI estimates for $16$-\QAM, varying additive SNR in dB and $\sigma_\Phi^2$ from $0$ to $0.1$.}
	\label{fig:CM_Capacity_MIKnuth_16QAM_APWGN}
\end{figure}

To verify the accuracy of the method, we compare the \MI estimate obtained with the presented method to the \MI of $16$-\QAM for a \PCAWGN channel.
The \MI is numerically calculated for a given modulation format and noise variances as described in \cite{Ungerboeck1982}.
\MI and its respective estimate are depicted in Fig.~\ref{fig:CM_Capacity_MIKnuth_16QAM_APWGN} for $16$-\QAM, varying \SNR values of the additive noise component $N$ of \eqref{eq:PCAWGN} from $-5$~dB to $25$~dB and a phase noise variance $\sigma_{\Phi}^2$ between $0$ (\cs \AWGN) and $0.1$. Higher values of $\sigma_{\Phi}^2$ were omitted because of their practical insignificance. The \MI estimate shown in Fig.~\ref{fig:CM_Capacity_MIKnuth_16QAM_APWGN} is very accurate over all simulated additive SNRs and for all values of $\sigma_{\Phi}^2$. For a very small overall noise contribution, the curves are even hardly distinguishable, showing an excellent match between our estimate and the true \MI. Further simulations have shown that similar accuracy is obtained for rectangular \QAM with $M$ up to $64$.

We conclude that the \MI estimation method is sufficiently accurate to determine the \MI for rectangular \QAM and transmission over a \ac{PCAWGN} channel. As this channel is a valid approximation for an optical fiber channel, it is justified to use our method to obtain an estimate of the \MI of an optical communication system.
%
%
%
%
\newpage
\section{Fiber Simulations}\label{sec:simulations}
In this section, we describe the simulation setup of an optical communication system. The transmission of \QAM symbols is simulated and \MI is estimated with the method presented in Section~\ref{sec:estimatingMI}.
\subsection{Simulation Setup}\label{subsec:simsetup}
\begin{figure*}[t]%
\centering
	\includegraphics[width=\columnwidth]{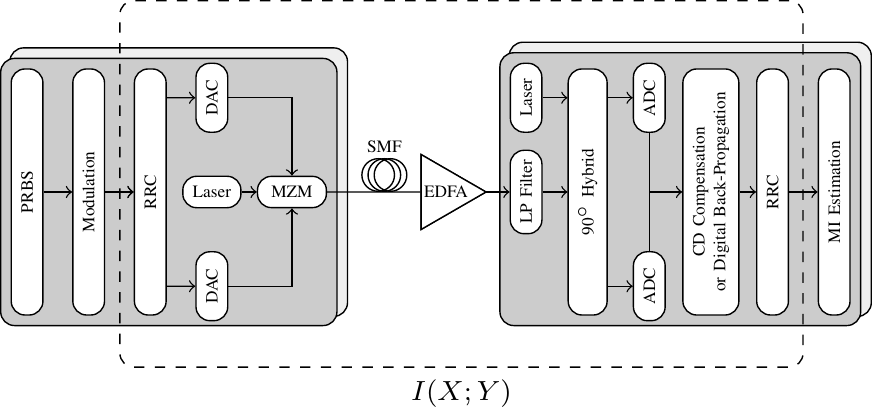}%
	\caption{Simulation setup of the examined optical communication system. The dashed box includes all components and subsystems that influence $I(X;Y)$.}
	\label{fig:simsetup}
\end{figure*}

A block diagram of the simulation setup is depicted in Fig.~\ref{fig:simsetup}. The default transmission and fiber link parameters are given in Tables~\ref{table:tx_params} and \ref{table:fiber_param}, respectively. Note that many of the components are considered ideal as the focus of this work is not to evaluate component imperfections but rather to estimate {the maximum achievable rate}, i.e. the \MI, of a coherent fiber system with transoceanic link length.

The leftmost block in Fig.~\ref{fig:simsetup} shows the \PDM transmitter. After generating a \PRBS, the bits are modulated. The resulting symbols are pulse-shaped in the digital domain with a \RRC filter with a roll-off factor of $0.05$ and converted into the analog domain with an ideal \DAC. The analog signal is transferred into the optical domain using an ideal push-pull \MZM with an ideal laser. The number of co-propagating \WDM channels, which are decorrelated copies of the central channel, is varied with the \WDM spacing such that the overall occupied bandwidth is as close as possible to $9\cdot50=450$~GHz while keeping the spectrum symmetrical around the center channel. This is to ensure that the channel of interest experiences approximately the same amount of nonlinearities during propagation, independent from the number of simulated channels. The relationship between channel spacing and number is shown vertically in the last two rows of Table~\ref{table:tx_params}. The above steps are repeated to generate the signal of the y-polarization. Lastly, the two polarizations are added and the signal is launched into the fiber.

The dual-polarization signal propagates over $60$ spans of \SMF, each followed by an \EDFA. The total transmission distance is $6000$~km with no inline dispersion compensation. Fiber propagation is simulated by solving the nonlinear Schrödinger equation numerically using the \SSFM \cite{Sinkin2003} with $32$~\SpS.

The rightmost block in Fig.~\ref{fig:simsetup} shows the coherent receiver. The central \WDM channel is extracted with an ideal band-pass filter whose bandwidth equals the \WDM spacing. The incoming optical signal is mixed with a local oscillator laser with zero linewidth and no frequency offset in a $90^{\circ}$ optical hybrid such that the in-phase and quadrature component of both polarizations are each made available to an ideal \ADC and then jointly to the \DSP subroutines with $32$~\SpS. The \DSP consists of ideal compensation of all \CD that is accumulated during propagation. For our simulation layout, equalization to counteract intersymbol interference is not required because there is no \PMD on the fiber and the Nyquist criterion is fulfilled by the \RRC filters at transmitter and receiver. For one set of simulations, ideal single-channel \DBP is performed, undoing both linear and nonlinear fiber effects. \CPR is not required because the lasers on the transmitter and receiver side are ideal and do not create phase perturbations. After matched \RRC filtering and downsampling, the \MI is calculated per polarization as outlined in Section~\ref{subsec:MIestimate} and then averaged over both polarizations.

The dashed box surrounding most of the simulation blocks in Fig.~\ref{fig:simsetup} depicts the channel for which the \MI is determined. Note that not only the fiber itself is part of the channel but also components and algorithms on both the transmitter and receiver side. Hence, the influence of these algorithms has an impact on the \MI and thus, on the net achievable data rate of the communication system.
\begin{table}[!b]
	\renewcommand{\arraystretch}{1.3}
	\caption{Transmission Parameters}
	\label{table:tx_params}
	\centering
	\begin{tabular}{|c||c|c|c|c|c|c|}
		\hline
		PRBS & \multicolumn{6}{c|}{$2^{18}$ bits per polarization} \\ \hline
		Modulation & \multicolumn{6}{c|}{\{$4$, $16$, $64$\}-\QAM}  \\ \hline
		\PDM & \multicolumn{6}{c|}{yes (independently)} \\ \hline
		Baud rate & \multicolumn{6}{c|}{$28$~GBaud} \\ \hline
		Data rate w/o \FEC & \multicolumn{6}{c|}{\{$112$, $224$, $336$\} Gbit/s}\\ \hline
		Pulse shaping & \multicolumn{6}{c|}{\RRC} \\ \hline
		Roll-off & \multicolumn{6}{c|}{$0.05$} \\ \hline
		\DAC/\ADC resolution & \multicolumn{6}{c|}{$\infty$ (ideal)} \\ \hline
		Laser linewidth & \multicolumn{6}{c|}{$0$ (ideal laser)} \\ \hline
		\MZM & \multicolumn{6}{c|}{ideal push-pull} \\ \hline
		Power per channel & \multicolumn{6}{c|}{$-8$ dBm to $4$ dBm}\\ \hline
		\WDM spacing [GHz] & $50$& $45$& $40$& $35$& $30$& $27.5$\\ \hline
		\WDM channels & $9$& $11$& $11$& $13$& $15$& $17$ \\ \hline
	\end{tabular}
	\vspace{10pt}
	\caption{Fiber Parameters}
	\label{table:fiber_param}
	\centering
	\begin{tabular}{|c||c|}
		\hline
		Fiber type & \SMF \\ \hline
		Carrier wavelength & $1550$~nm \\ \hline
		Length per span & $100$~km \\ \hline
		Number of spans & $60$ \\ \hline
		Attenuation $\alpha$ & $0.2\frac{\textnormal{dB}}{\textnormal{km}}$ \\ \hline
		Nonlinear coefficient  $\gamma$ & $1.3\frac{\textnormal{1}}{\textnormal{W}\cdot\textnormal{km}}$ \\ \hline
		Chromatic dispersion & $17\frac{\textnormal{ps}}{\textnormal{nm}\cdot\textnormal{km}}$ \\ \hline
		Dispersion slope & $0$ \\ \hline
		\PMD & $0$ \\ \hline
		SSFM step size & $0.1$~km \\ \hline
		\SpS & $32$ \\ \hline
		EDFA noise figure & $4$~dB \\ \hline
	\end{tabular}
\end{table}%
\subsection{Simulation Results}\label{subsec:simresults}
For the outlined simulation setup, \QAM of order $4$, $16$, and $64$ is compared in terms of the respective \MI. Additionally, the impact of ideal single-channel \DBP is examined for $16$-\QAM and $64$-\QAM, and the \SE is analyzed for different \WDM spacings.
\subsubsection{Different modulation schemes}
\begin{figure}[t]%
	\includegraphics[width=\figurewidth,height=\figureheight]{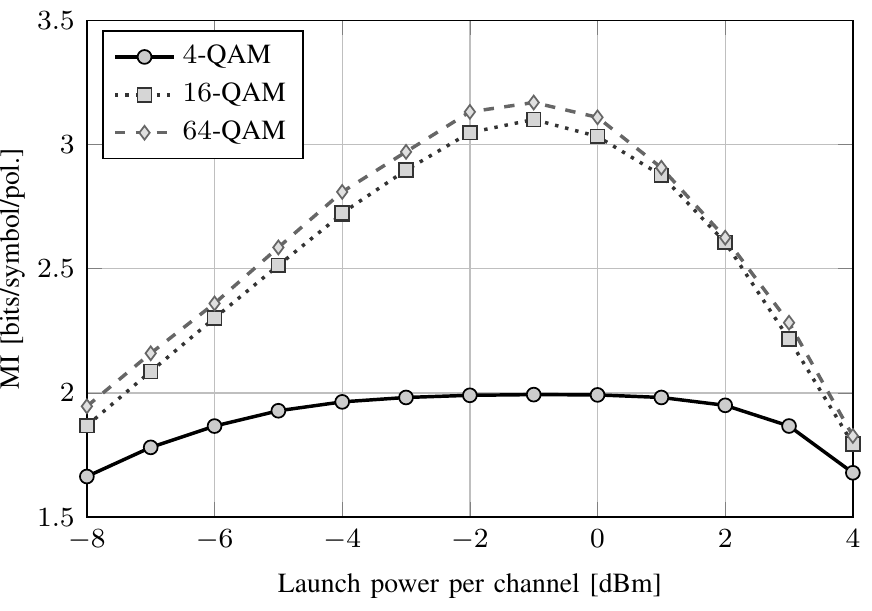}%
	\caption{{\MI} over launch power per channel for \QAM of order $4$, $16$, $64$.}
	\label{fig:MIoverPower_6000km_QPSK_16QAM_64QAM}
\end{figure}

In Fig.~\ref{fig:MIoverPower_6000km_QPSK_16QAM_64QAM}, the \MI per polarization is depicted over the launch power per channel in increments of $1$~dBm. The curve for each modulation scheme follows the  behavior that is typical for a nonlinear fiber system. In the low-power regime, the performance is limited by noise originating from the optical amplifiers. The maximum \MI is obtained at a power of $-1$~dBm for all modulations. This is in agreement with \cite{Kikuchi2011} where it is analytically shown that the optimum launch power is constant for rectangular \QAM of identical symbol rate. Beyond $-1$~dBm, the impact of the nonlinearities becomes larger and reduces \MI. Previous results obtained by considering an \HD code and \BERin as figure of merit show a similar shape \cite{Kilmurray2012}.
We will now discuss in more detail the maximum \MI of $3.1$~bits/symbol for $16$-\QAM obtained at the optimum launch power of $-1$~dBm. This equals a net channel data rate of $174$~Gbit/s in comparison to $100$~Gbit/s of commercially available link based on \QPSK.\\
We know from Section~\ref{subsec:mi} that \MI {is the achievable rate of transmission}. Out of every $\log_2(M)=4$ bits per $16$-\QAM symbol, up to $3.1$ bits can carry information and at least $0.9$ bits must be reserved for the redundancy of the assumed ideal coding scheme. The maximum rate $R$ that can be used while still achieving an arbitrarily small \BER equals $\frac{3.1}{\log_2(16)} \simeq 0.78$. Consequently, any code with $R>0.78$ will inevitably lead to a significant, i.e. not arbitrarily small \BERout if the channel memory is neglected. An identical line of argument can be made for \QPSK and $64$-\QAM. At the optimum launch power, $R=\frac{1.99}{\log_2(4)} \simeq 0.99$ for \QPSK and $R=\frac{3.17}{\log_2(64)} \simeq 0.53$ for $64$-\QAM. Clearly, \QPSK is a suboptimal choice of modulation because it only achieves $1.99$~bits/symbol while more than one additional bit of information per symbol is available by using $16$-\QAM. Fig.~\ref{fig:Capacities_AWGN} also provides a hint that a higher modulation than \QPSK should be used for the considered long-haul link. Assume that the optical system operates at the optimal launch power of $-1$~dBm. The $1.99$~bits/symbol achieved by \QPSK at this power correspond to about $10$~dB \SNR for a hypothetical equivalent \AWGN channel. At $10$~dB SNR, we could get to $3.16$~bits/symbol by using $16$-\QAM and still have an arbitrarily small \BER. Consequently, we prefer $16$-\QAM to \QPSK for an \AWGN channel as well as for the outlined optical fiber system with nonlinearities.

Comparing the two highest-order modulation schemes in Fig.~\ref{fig:MIoverPower_6000km_QPSK_16QAM_64QAM}, $64$-\QAM slightly outperforms $16$-\QAM over the entire range of launch powers. At the optimum launch power, the respective \ac{MI}s differ by less than $0.1$~bits/symbol. In order to make use of this little gain, a channel code of smaller rate is required for higher modulations. Also, the increased number of constellation levels imposes higher requirements on components such as the \ADC and \DAC or on the laser linewidth, and also on \DSP algorithms such as \CPR. We argue that for the simulation setup of Section~\ref{subsec:simsetup} with a transoceanic \SMF link, $16$-\QAM is the modulation format that offers the best tradeoff between \MI and complexity. Resorting again to Fig.~\ref{fig:Capacities_AWGN} as an illustration and assuming that we operate in an equivalent \AWGN channel, there is only about $0.1$~bits/symbol to be gained by using $64$-\QAM instead of $16$-\QAM at an \SNR of $10$~dB. This supports our result that $16$-\QAM is the best choice for our long-haul simulation setup. {This is in agreement with experimental results} \cite{Djordjevic2010} where for $5000$~km \SMF and higher \EDFA noise figures, using a modulation format with more than $16$~states does not result in a notable gain.

One might argue that four-dimensional (4D) modulations based on set-partitioning (SP) \cite{Agrell2009} have shown better performance than modulating each polarization independently. Experiments \cite{Eriksson2013} suggest that SP-$128$-QAM \cite{Coelho2011} can outperform \DP-$16$-\QAM by $1.9$~dB SNR at a \BERin of $10\textsuperscript{-3}$ and identical net bit rate. This comparison, however, is biased because SP-$128$-\QAM is essentially \DP-$16$-\QAM with an inherent \SPC code of rate $R=\frac{\log_2(M)-1}{\log_2(M)}=\frac{7}{8}$ applied such that the Euclidean distance between neighboring symbols in the 4D constellation space is increased. With such a coding scheme in place, SP-$128$-\QAM clearly must be better than uncoded \DP-$16$-\QAM. However, \SPC is not ideal in terms of coding performance but its strength rather lies in the very low complexity. 
Accordingly, better performance is achieved if an ideal code of rate $\frac{7}{8}$ instead of the \SPC is used. SP-$128$-QAM as \DP-$16$-\QAM with an inherent low-performance low-complexity code is hence not treated in this work as we focus on the maximum achievable rate. The same argument holds for schemes of different orders that are also based on SP.
\begin{figure}[t]%
	\includegraphics[width=\figurewidth,height=\figureheight]{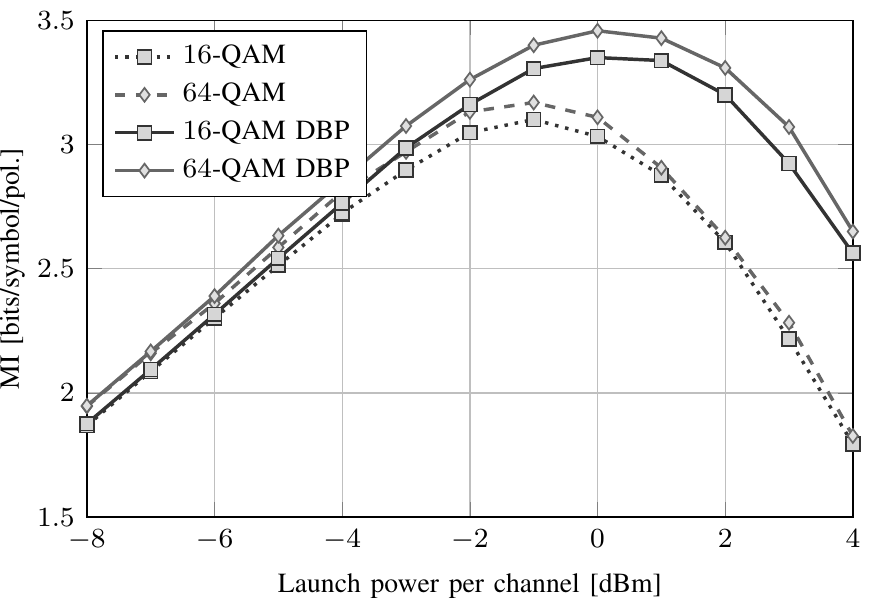}%
	\caption{The effect of ideal single-channel \DBP for $16$-\QAM and $64$-\QAM after $6000$~km of transmission.}
	\label{fig:MIoverPower_16QAM_DBP_6000km}
\end{figure}%
\subsubsection{Digital back-propagation}
In the following, \DBP is analyzed as an example of the impact of a particular \DSP algorithm on the achievable rate. \DBP is a powerful yet complex approach to mitigate the nonlinear signal interactions by back-propagating the signal in the digital domain at the receiver \cite[Sec.~2]{Li2008}. We perform ideal single-channel \DBP with a step size that is identical to the one used for the \SSFM and also with identical \SpS. The results are shown in Fig.~\ref{fig:MIoverPower_16QAM_DBP_6000km} for \QAM of order $16$ and $64$, and a \WDM spacing of $50$~GHz. All other parameters are as presented in Section~\ref{subsec:simsetup}. The maximum \MI is obtained at $0$~dBm launch power instead of -1~dBm without \DBP as the amount of nonlinearities is effectively reduced by \DBP. The maximum \MI is $3.35$~bits/symbol for $16$-\QAM and $3.45$~bits/symbol for $64$-\QAM, which corresponds to an increase by $0.25$~bits/symbol and $0.28$~bits/symbol, respectively, compared to the case without \DBP. As both step size and \SpS are identical to the \SSFM, this is the maximum gain possible by deterministic single-channel \DBP for the presented simulation setup.%
\subsubsection{\WDM spacings}
\begin{figure}[!t]%
	\includegraphics[width=\figurewidth,height=\figureheight]{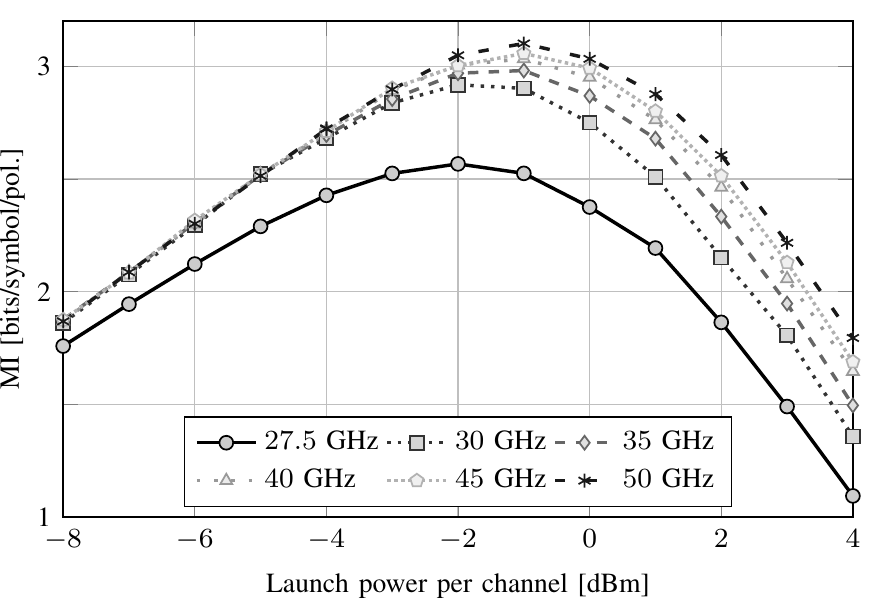}%
	\caption{{\MI} over power per channel for varying \WDM spacings and $16$-\QAM.}
	\label{fig:MIoverPtx_WDMspacing_6000km}
\end{figure}

In Fig.~\ref{fig:MIoverPtx_WDMspacing_6000km}, \MI is depicted for varying launch powers, \WDM spacings from $27.5$~GHz to $50$~GHz, and without \DBP. The number of co-propagating channels is varied according to Table~\ref{table:tx_params}.
Decreasing the spacing from $50$ GHz to $30$ GHz does not affect \MI in the linear regime. Once the launch power is increased such that the system is driven into the nonlinear regime, \MI declines with decreasing channel spacing. For $30$~GHz, $2.9$ bits/symbol are achievable. While this behavior is qualitatively expected as the phase mismatch between co-propagating signals increases with spectral spacing and hence the impact of four-wave mixing decreases \cite[Sec.~10.1]{Agrawal2007}, it is interesting to consider it quantitatively. A decrease in channel spacing by $60$\% leads to a drop in \MI of only about $6.5\%$, from $3.1$~bits/symbol to $2.9$~bits/symbol.\\
If the channel spacing is further reduced to $27.5$ GHz, the \MI significantly decreases not only for high launch powers but also in the linear regime. The channel spacing is in this case smaller than the $28 \cdot 1.05 = 29.4$~GHz occupied by the signal. As the interference due to this spectral overlap is essentially noise, the optical signal is subject to additional impairments and the \MI decreases.

\begin{figure}[!t]%
	\includegraphics[width=\figurewidth,height=\figureheight]{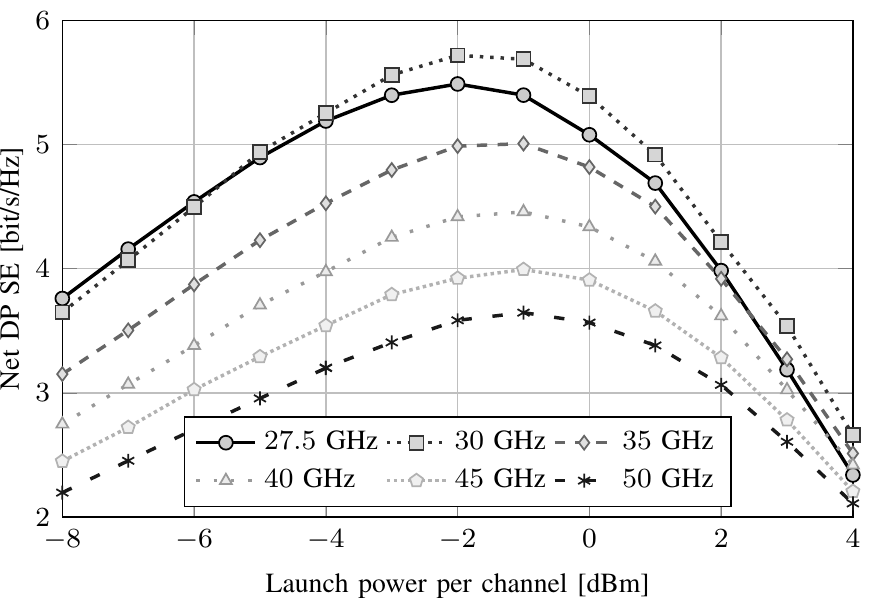}%
	\caption{Net \DP \SE over power per channel for varying \WDM spacings and $16$-\QAM.}
	\label{fig:SEoverPtx_WDMspacing_6000km}
\end{figure}

Most of the \MI curves in Fig.~\ref{fig:MIoverPtx_WDMspacing_6000km} are close to each other, yet the channel spacings differ a lot. This leads to investigate \SE, shown in Fig.~\ref{fig:SEoverPtx_WDMspacing_6000km}. Note that net \SE is depicted, i.e. coding overhead for an ideal \FEC scheme is taken into account and an arbitrarily small \BER is achievable by such a code.
%
A maximum \DP \SE of $5.7$~bits/s/Hz is obtained for a \WDM spacing of $30$~GHz. For this spacing almost the entire available spectrum is used, which more than compensates for the slightly smaller \MI found in Fig.~\ref{fig:MIoverPtx_WDMspacing_6000km}.
To put this \SE into perspective, it is by $0.7$~bit/s/Hz larger than the $5$~bits/s/Hz \SE that has been achieved experimentally over $6600$~km pure-silica fiber core fiber and a hybrid \ac{EDFA}-Raman amplification scheme \cite{Salsi2013}.

\SE in general decreases with increasing channel spacing as long as the \WDM spacing is larger than the bandwidth occupied by the signal. Somewhat surprisingly, $27.5$~GHz gives the highest \SE in the linear regime up to $-4$~dBm and performs well also for high launch powers. This behavior, however, is expected to vanish if an optical network with add-drop multiplexers is considered. The repeated filtering in combination with the spectral broadening of the pulse will further deteriorate the \MI for a fiber system with sub-Nyquist \WDM spacing.
%

\section{Conclusion}\label{sec:conclusion}
In this work, we estimate \MI by means of histograms for which a method to find the correct bin number is presented.
%
%
A simplified channel model for the optical fiber channel is used to show the accuracy of our \MI estimation method. The method is applied to a coherent optical fiber system with $6000$~km of standard fiber and the \MI estimate is used as both an absolute and relative figure of merit. For ideal transmitter and receiver components yet without \DBP, $16$-\QAM proves to be the modulation format that offers the best tradeoff between complexity and data rate {for transoceanic distances}. An \MI of $3.1$~bits/symbol is achievable for $16$-\QAM and a $50$~GHz spacing. This mean that any \FEC for the considered system must have an overhead of at least $29$\% if memory from the channel is not considered. The amount of memory in an optical fiber system and how to make use of it remain open interesting questions as they may lead to a further increase in data rate.\\ 
\MI also allows to relatively quantify the impact of receiver subsystems. Using ideal single-channel \DBP as an example, a respective maximum gain of $0.25$~bits/symbol for $16$-\QAM and $0.28$~bits/symbol for $64$-\QAM is achieved with only the center channel back-propagated. By extending this to more channels, the impact of nonlinear interchannel interactions and also the impact of nonlinear signal-noise interactions can be quantified. When the \WDM channel spacing is decreased to almost the signal bandwidth, a maximum \SE of $5.7$~bits/s/Hz over both polarizations is possible. While this value can be further increased by employing a scheme to mitigate fiber nonlinearities, it is questionable whether the potential increase in data rate is sufficient to cater for future capacity requirements.\\
The presented analysis is applicable to any component or method employed between the modulator on the transmitter side and the decoder input at the receiver in order to quantify the potential data rate gain or loss of a subsystem. For example, data-aided equalization can be compared to blind equalization as to the achievable rate decreased by the training symbols. Also, the effect of upgrading to more advanced amplification schemes or fibers can be quantified. 
%
\bibliographystyle{IEEEtran}
\input{references.bbl}

%
%
%
\end{document}

%% file: packages_paper.tex
\usepackage[printonlyused]{acronym}

\usepackage{stfloats}
\usepackage{xspace}
\usepackage{ifthen}
\usetikzlibrary{spy}

%% file: myAcronyms.tex
\acrodef{MI}{mutual information}
\newcommand{\MI}{\ac{MI}\xspace}
\acrodef{BER}{bit error ratio}
\newcommand{\BER}{\ac{BER}\xspace}
\acrodef{AWGN}{additive white Gaussian noise}
\newcommand{\AWGN}{\ac{AWGN}\xspace}
\acrodef{cs}[c.~s.]{circularly symmetric}
\newcommand{\cs}{\ac{cs}\xspace}
\acrodef{PCAWGN}[p.c. AWGN]{partially coherent AWGN}
\newcommand{\PCAWGN}{\ac{PCAWGN}\xspace}
\acrodef{pdf}{probability density function}
\newcommand{\pdf}{\ac{pdf}\xspace}
\acrodef{pmf}{probability mass function}
\newcommand{\pmf}{\ac{pmf}\xspace}
\acrodef{CM}{coded modulation}
\acrodef{BICM}{bit-interleaved coded modulation}
\acrodef{QAM}{quadrature amplitude modulation}
\newcommand{\QAM}{\ac{QAM}\xspace}
\acrodef{SNR}{signal-to-noise ratio}
\newcommand{\SNR}{\ac{SNR}\xspace}
\acrodef{DSP}{digital signal processing}
\newcommand{\DSP}{\ac{DSP}\xspace}
\acrodef{QPSK}{quadrature phase-shift keying}
\newcommand{\QPSK}{\ac{QPSK}\xspace}
\acrodef{FEC}{forward error correction}
\newcommand{\FEC}{\ac{FEC}\xspace}
\acrodef{BPSK}{binary phase-shift keying}

\acrodef{PDM}{polarization-division multiplexed}
\newcommand{\PDM}{\ac{PDM}\xspace}
\acrodef{RS}{Reed-Solomon}
\newcommand{\RS}{\ac{RS}\xspace}
\acrodef{LDPC}{low-density parity-check}
\newcommand{\LDPC}{\ac{LDPC}\xspace}
\acrodef{RV}{random variable}

\acrodef{WDM}{wavelength-division multiplexed}
\newcommand{\WDM}{\ac{WDM}\xspace}
\acrodef{HD}{hard-decision}
\newcommand{\HD}{\ac{HD}\xspace}
\acrodef{SD}{soft-decision}
\newcommand{\SD}{\ac{SD}\xspace}
\acrodef{RRC}{root-raised cosine}
\newcommand{\RRC}{\ac{RRC}\xspace}
\acrodef{DAC}{digital-to-analog converter}
\newcommand{\DAC}{\ac{DAC}\xspace}
\acrodef{ADC}{analog-to-digital converter}
\newcommand{\ADC}{\ac{ADC}\xspace}
\acrodef{MZM}{Mach-Zehnder modulator}
\newcommand{\MZM}{\ac{MZM}\xspace}
\acrodef{SMF}{single-mode fiber}
\newcommand{\SMF}{\ac{SMF}\xspace}
\acrodef{CD}{chromatic dispersion}
\newcommand{\CD}{\ac{CD}\xspace}
\acrodef{EDFA}{Erbium-doped fiber amplifier}
\newcommand{\EDFA}{\ac{EDFA}\xspace}
\acrodef{LO}{local oscillator}

\acrodef{LP}{low-pass}

\acrodef{CPR}{Carrier phase recovery}
\newcommand{\CPR}{\ac{CPR}\xspace}
\acrodef{DBP}{digital back-propagation}
\newcommand{\DBP}{\ac{DBP}\xspace}
\acrodef{PRBS}{pseudo-random binary sequence}
\newcommand{\PRBS}{\ac{PRBS}\xspace}
\acrodef{PMD}{polarization-mode dispersion}
\newcommand{\PMD}{\ac{PMD}\xspace}
\acrodef{SSFM}{split-step Fourier method}
\newcommand{\SSFM}{\ac{SSFM}\xspace}
\acrodef{SpS}{samples per symbol}
\newcommand{\SpS}{\ac{SpS}\xspace}
\acrodef{ISI}{inter-symbol interference}

\acrodef{SE}{spectral efficiency}
\newcommand{\SE}{\ac{SE}\xspace}
\acrodef{SPQAM}{set-partitioning QAM of order}

\acrodef{SPC}{single parity check}
\newcommand{\SPC}{\ac{SPC}\xspace}
\acrodef{DP}{dual-polarization}
\newcommand{\DP}{\ac{DP}\xspace}
\newcommand{\Q}{\ensuremath{Q^2}\xspace}
\newcommand{\BERin}{\ensuremath{\text{BER}_{\text{in}}}\xspace}
\newcommand{\BERout}{\ensuremath{\text{BER}_{\text{out}}}\xspace}

\DeclareMathOperator*{\argmax}{arg\,max}